\documentclass[%
reprint,
showpacs, 
preprintnumbers,
amsmath,
amssymb,
twocolumn,
prl,
graphicx,
]{revtex4-1}

\usepackage{mathptmx}
\usepackage{graphicx}
 \usepackage{dcolumn}
\usepackage{bm}
\usepackage{braket}

\begin{document}

\title{Analog quantum error correction with encoding a qubit into an oscillator}

\author{Kosuke Fukui}
\author{Akihisa Tomita}%
\author{Atsushi Okamoto}%
\affiliation{%
Graduate School of Information Science and Technology, Hokkaido University \\ Kita14-Nishi9, Kita-ku, Sapporo 060-0814, Japan
}%

\date{\today}

\begin{abstract}
To implement fault-tolerant quantum computation with continuous variables, Gottesman--Kitaev--Preskill (GKP) qubits have been recognized as an important technological element. However, the analog outcome of GKP qubits, which includes beneficial information to improve the error tolerance, has been wasted, because the GKP qubits have been treated as only discrete variables. In this letter, we propose a hybrid quantum error correction approach that combines digital information with the analog information of the GKP qubits using a maximum-likelihood method. As an example, we demonstrate that the three-qubit bit-flip code can correct double errors, whereas the conventional method based on majority voting on the binary measurement outcome can correct only a single error. As another example, we show that a concatenated code known as Knill's $C_{4}/C_{6}$ code can achieve the hashing bound for the quantum capacity of the Gaussian quantum channel (GQC). To the best of our knowledge, this approach is the first attempt to draw both digital and analog information to improve quantum error correction performance and achieve the hashing bound for the quantum capacity of the GQC.
\end{abstract}

\maketitle


Quantum computation (QC) has a great deal of potential~\cite{Shor, Grov}. Although small-scale quantum circuits with various qubits have been demonstrated~\cite{Niem, Blat}, a large-scale quantum circuit that requires scalable entangled states is still a significant experimental challenge for most candidates of qubits. In continuous variable (CV) QC, squeezed vacuum (SV) states with the optical setting have shown great potential to generate scalable entangled states because the entanglement is generated by only beam splitter (BS) coupling between two SV states~\cite{Yoshi}. However, scalable computation with SV states has been shown to be difficult to achieve because of the accumulation of errors during the QC process, even though the states are created with perfect experimental apparatus~\cite{Meni}. Therefore, fault-tolerant (FT) protection from noise is required that uses the quantum error correcting code. Because noise accumulation originates from the \textquotedblleft continuous\textquotedblright\, nature of the CVQC, it can be circumvented by encoding CVs into digitized variables using an appropriate code, such as Gottesman--Kitaev--Preskill (GKP) code~\cite{GKP}, which are referred to as GKP qubits in this letter. Menicucci showed that CV-FTQC is possible within the framework of measurement-based QC using SV states with GKP qubits~\cite{Meni}. Moreover, GKP qubits keep the advantage of SV states on optical implementation that they can be entangled by only BS coupling. Hence, GKP qubits offer a promising element for the implementation of CV-FTQC.

To be practical, the squeezing level required for FTQC should be experimentally achievable. Unfortunately, Menicucci's scheme still requires a 14.8 dB squeezing level to achieve the FT threshold $2 \times 10^{-2}$~\cite{Knill, Fuji1, Fuji2}. Thus, another twist is necessary to reduce the required squeezing level. It is analog information contained in the GKP qubit that has been overlooked. The effect of noise on CV states is observed as a deviation in an analog measurement outcome, which includes beneficial information for quantum error correction (QEC). Despite this, the analog information from the GKP qubit has been wasted because the GKP qubit has been treated as only a discrete variable (DV) qubit, for which the measurement outcomes are described by bits. Harnessing the wasted information for the QEC will improve the error tolerance compared with using the conventional method based on only bit information. Such a use of analog information has been developed in classical error correction against the disturbance such as an additive white Gaussian noise~\cite{Kai} and identified as an important tool for qubit readout~\cite{Dan,Dan2}. However, the use of analog information has been left unexploited to improve the QEC performance~\cite{Cap1}.

In this letter, we propose a maximum-likelihood method (MLM) using the analog outcome and demonstrate the advantage of our scheme using numerical simulations for two remarkable examples. First, we show that the three-qubit bit-flip code can correct double bit-flip errors effectively using our method, in contrast to the conventional method that uses DV information that can correct only a single error. Second, we show that the concatenated code with Calderbank--Shor--Steane codes, particularly the $C_{4}/C_{6}$ code proposed by Knill~\cite{Knill}, can achieve the hashing bound for the quantum capacity of the Gaussian quantum channel (GQC)~\cite{GKP,Harri}, which implies that our technique improves the GKP qubit into one of the optimal encoded states against the disturbance in the GQC.

{\it The GKP qubit}.---We review the GKP qubit and error model considered in this letter. Gottesman, Kitaev, and Preskill proposed a method to encode a qubit in an oscillator's $q$ (position) and $p$ (momentum) quadratures to correct errors caused by a small deviation in the $q$ and $p$ quadratures. The basis of the GKP qubit is composed of a series of Gaussian peaks of width $\sigma$ and separation $\sqrt{\pi}$ embedded in a larger Gaussian envelope of width 1/$\sigma$. Although in the case of infinite squeezing ($\sigma \rightarrow 0$) the GKP qubit bases become orthogonal, in the case of finite squeezing, the approximate code states are not orthogonal and there is a probability of misidentifying $\ket {\widetilde{0}}$ as $\ket {\widetilde{1}}$, and vice versa. Provided the measured magnitude deviates less than $\sqrt{\pi}/2$ from the peak value, the decision of the bit value from the measurement of the GKP qubit is correct. The probability $p_{corr}$ that we identify the correct bit value is the portion of a normalized Gaussian of a variance ${{\sigma}}^2$ that lies between $-\sqrt{\pi}/2$ and $\sqrt{\pi}/2$~\cite{Meni}:
\begin{equation}
p_{\rm corr} = \int_{\frac{-\sqrt{\pi}}{2}}^{\frac{\sqrt{\pi}}{2}} dx \frac{1}{\sqrt{2\pi {\sigma} ^2}} {\rm exp}(-x^2/{2{\sigma} ^2}).
\end{equation}
In addition to the imperfection that originates from the finite squeezing of the initial states, we consider the GQC~\cite{GKP,Harri}, which leads to a displacement in the quadrature during the QC process. The channel is described by superoperator $\zeta $ acting on density operator $\rho$ as follows: 
\begin{equation}
\rho \to \zeta (\rho) = \frac{1}{\pi{\xi }^2}\int d^2\alpha \mathrm{e}^{-{| \alpha |}^2/{{\xi}^2}}D( \alpha ) \rho D( \alpha ) ^{\dagger },
\end{equation}
where $D(\alpha)$ is a displacement operator in the phase space. The position $q$ and momentum $p$ are displaced independently as follows:
\begin{equation}
 q \to q + v , \  \ p \to p + u,
 \end{equation}
where $v$ and $u$ are real Gaussian random variables with mean zero and variance $\xi ^2$. Therefore, the GQC conserves the position of the Gaussian peaks in the probability density function on the measurement outcome of the GKP qubit, but increases the variance as $\xi ^2$.

\begin{figure}[t]
 \includegraphics[angle=270, width=1.0\columnwidth]{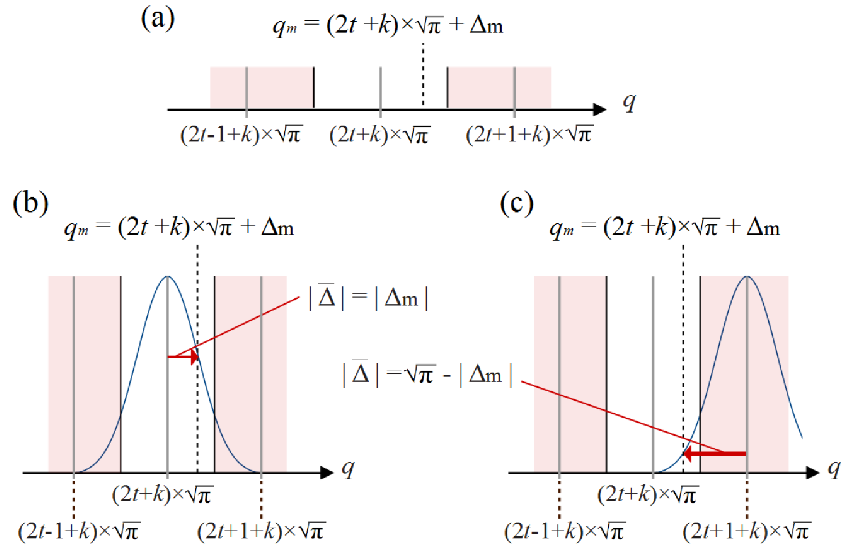}
 \caption{\label{figa}Introduction of a likelihood function. (a) Measurement outcome and deviation from the peak value in $q$ quadrature. The dotted line shows the measurement outcome $q_{\rm m}$ equal to $(2 t + k)\sqrt{\pi}+\Delta_{\rm m}$ $(t = 0, \pm 1, \pm 2,\cdots,\ k = 0, 1)$, where $k$ is defined as the bit value that minimizes the deviation $\Delta_{\rm m}$. The red areas indicate the area that yields code word $(k+1)$ mod 2, whereas the white area denotes the area that yields the codeword $k$. (b) and (c) Gaussian distribution functions as likelihood functions of the true deviation value $\overline{\Delta}$ represented by the arrows. (b) refers to the case of the correct decision, where the amplitude of the true deviation value is $|\overline{\Delta}| < \sqrt{\pi}/2$, whereas (c)  the case of the incorrect decision $ \sqrt{\pi}/2 < |\overline{\Delta}| < \sqrt{\pi}$.}
\end{figure}

{\it Likelihood function}.---We make a decision on the bit value $k ( = 0, 1)$ from the measurement outcome of the GKP qubit $q_{\rm m}= q_{k}+ \Delta_{\rm m}$ to minimize the deviation $|\Delta_{\rm m}|$, where $q_{k}(k = 0,1)$ is defined as $(2 t + k)\sqrt{\pi} (t = 0, \pm 1, \pm 2,\cdots.)$, shown in Fig.\ref{figa}(a). If we consider only digital information $k$, as in conventional QEC, we waste the analog information contained in $\Delta_{\rm m}$. 

Instead, we propose a likelihood method to improve our decision for the QEC using analog information. We define the true deviation $|\bar{\Delta}|$ as the difference between the measurement outcome and true peak value $\bar{q}_k$, that is, $|\bar{\Delta}|=|\bar{q}_k-q_m|$. We consider the following two possible events: one is the correct decision, where the true deviation value $|\overline{\Delta }|$ is less than $\sqrt{\pi}/2$ and equals to $|\Delta_{\rm m}|$ as shown in Fig.\ref{figa}(b). The other is the incorrect decision, where $|\overline{\Delta}|$ is greater than $\sqrt{\pi}/2$ and satisfies $|\bar{\Delta}|+|\Delta_m|=\sqrt{\pi}$, as shown in Fig.\ref{figa}(c). Because the true deviation value obeys the Gaussian distribution function $f(\overline{\Delta})$, we can evaluate the probabilities of the two events by
\begin{eqnarray}
f(\overline{\Delta}) = \frac{1}{\sqrt{2\pi\sigma^{2}}} \mathrm{e}^{-\overline{\Delta}^{2}/(2\sigma^{2})}.
\end{eqnarray}
In our method, we regard function $f (\overline{\Delta})$ as a likelihood function. Using this function, the likelihood of the correct decision is calculated by $f(\overline{\Delta}) = f(\Delta_{\rm m})$. The likelihood of the incorrect decision, whose $|\overline{\Delta}|$ is $\sqrt{\pi}-|\Delta_{\rm m}|$, is calculated by $f(\overline{\Delta}) =f(\sqrt{\pi}-|\Delta_{\rm m}|)$. We can reduce the decision error on the entire code word by considering the likelihood of the joint event and choosing the most likely candidate. 

{\it Bit--flip code with analog information}.---To provide an insight into our method, we focus on the three-qubit bit-flip code as a simple example. In this code, a single logical qubit $\ket{\widetilde{\psi}}_{\rm L}$=$\alpha\ket{\widetilde{0}}_{\rm L}+ \beta\ket{\widetilde{1}}_{\rm L}$, where ${|\alpha |}^{2} + {|\beta |}^{2}=1 $, is encoded into three GKP qubits. The two logical basis states $\ket{\widetilde{0}}_{{\rm L}}$ and $\ket{\widetilde{1}}_{{\rm L}}$ are defined as $\ket{\widetilde{0}}_{{\rm L}}=\ket{\widetilde{0}}_{{\rm 1}}\ket{\widetilde{0}}_{{\rm 2}}\ket{\widetilde{0}}_{{\rm 3}}$ and $\ket{\widetilde{1}}_{{\rm L}}=\ket{\widetilde{1}}_{{\rm1}}\ket{\widetilde{1}}_{{\rm 2}}\ket{\widetilde{1}}_{{\rm 3}}$, respectively. 

In the QEC with the three-qubit bit-flip code, the error identification for the GKP qubits is substantially different from that for DV-QEC. While the parity of the code qubits is transcribed on the ancilla qubit in DV-QEC, the deviation of the physical GKP qubits is projected onto the deviation of the ancillae (see the supplementary information for the details). From the measurement of the three ancillae in $q$ quadrature, we obtain the outcome $q_{{\rm m, A}i} = q_{0}+ \Delta_{{\rm m, A}i}$ ($i$ = 1, 2) from ancillae 1 and 2, and $q_{{\rm m, A}3} = q_{k}+ \Delta_{{\rm m, A}3}$ ($k$ = 0, 1) from ancilla 3, under the conditions $\Delta_{{\rm m, A}i}\in[-\sqrt{\pi},\sqrt{\pi}]$ and $\Delta_{{\rm m, A}3}\in[-\sqrt{\pi}/2,\sqrt{\pi}/2 ]$. We then define the values $\delta_{1}$ = $\Delta_{{\rm m, A}1}-\Delta_{{\rm m, A}2}+\Delta_{{\rm m, A}3}$ and $\delta_{2}$ = $\Delta_{{\rm m, A}2}-\Delta_{{\rm m, A}3}$. For $i$ = 1, 2, if $\delta_{i}\in[-\sqrt{\pi},\sqrt{\pi}]$, then we define the values $M_{i}$=$\delta_{i}$. Otherwise, if $\delta_{i}\in[\sqrt{\pi},2\sqrt{\pi}]$, we define the values $M_{i}$=$\delta_{i}-2\sqrt{\pi}$, and if  $\delta_{i}\in[-2\sqrt{\pi},-\sqrt{\pi}]$, we define the values $M_{i}$= $2\sqrt{\pi}+\delta_{i}$. Error identification is executed from $M_{1}$ and $M_{2}$ as follows. If both $|M_{1}|$ and $|M_{2}|$ are smaller than $\sqrt{\pi}/2$, we decide that no error occurs on the logical qubits. Otherwise, we consider two error patterns: one containing a single error, and the other containing double errors. For the first pattern, we presume that the true deviation values $\overline{\Delta}_{i}$ ($i$ = 1, 2) and $\overline{\Delta}_{\rm 3}$ of the qubits in the logical qubit are $M_{i}$ and $\Delta_{{\rm m, A}3}$, respectively. Then, the likelihood of the first pattern $F_{1}$ is given by $F_{1} = f(M_{1})f(M_{2})f({\Delta}_{\rm m, A3})$. For the second pattern, if $M_{i}\in[0,\sqrt{\pi}]$, we presume that $\overline{\Delta}_{i}$ is ${M_{i}}^{\ast }=M_{i}-\sqrt{\pi}$, and if $M_{i}\in[-\sqrt{\pi},0]$, we presume that $\overline{\Delta}_{i}$ is ${M_{i}}^{\ast }=M_{i}+\sqrt{\pi}$. If $\Delta_{{\rm m, A}3}\in [0,\sqrt{\pi}/2]$, we presume $\overline{\Delta}_{3}$ to be ${\Delta}_{{\rm m, A}3}^{\ast }=\Delta_{{\rm m, A}3}-\sqrt{\pi}$, and if $\Delta_{{\rm m, A}3}\in[-\sqrt{\pi}/2,0]$, we presume that $\overline{\Delta}_{3}$ is ${\Delta}_{{\rm m, A}3}^{\ast }=\Delta_{{\rm m, A}3}+\sqrt{\pi}$. Then, the likelihood of the second pattern $F_{2}$ is given by $F_{2} = f({M_{1}}^{\ast })f({M_{2}}^{\ast })f({\Delta}_{{\rm m, A}3}^{\ast })$. Hence, we can use the likelihood functions $f(|\Delta_{\rm m}|)$ and $f(\sqrt{\pi}-|\Delta_{\rm m}|)$ to compare the two error patterns and decide the more likely pattern. For example, if $M_{1}$ is in the range $[\sqrt{\pi}/2,\sqrt{\pi}]$, and both $M_{2}$ and $\Delta_{{\rm m, A}3}$ are in the range $[0,\sqrt{\pi}/2]$, we consider the first error pattern as a single error on qubit 1 of the logical qubit and the second error pattern as double errors on qubits 2 and 3. If $F_{\rm 1} >  F_{\rm 2}$, we decide that the first error pattern occurs, and vice versa. In error identification, the likelihood that $|\overline{\Delta}_{i}|$ is greater than $\sqrt{\pi}$ is not taken into account because it is always less than $\sqrt{\pi}$ provided $|\overline{\Delta}_{i}|$ is less than $\sqrt{\pi} $. In the conventional manner, based on majority voting with binary measurement outcomes, the first error pattern is invariably selected because an estimation using only digital information yields a larger probability for a single error than that for double errors.

\begin{figure}
 \includegraphics[angle=270, width=1.05\columnwidth]{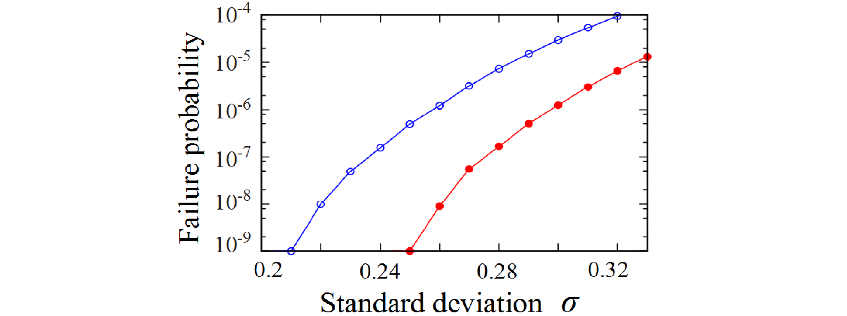}
 \caption{\label{figb}Simulation results for the failure probabilities of the three-qubit bit-flip code using the conventional (blue line with open circles) and proposed methods (red line with filled circles).}
\end{figure}

We numerically simulated the QEC for the three-qubit bit-flip code using the Monte Carlo method. In this simulation, it is assumed that the encoded data qubit is prepared perfectly, that is, the initial variances of the data qubit and ancillae are zero, and the variances of the GKP qubits of the encoded data qubit increase independently in the GQC. These assumptions are set to allow a clear comparison between the conventional and proposed methods. In Fig.\ref{figb}, the failure probabilities of the QEC are plotted as a function of the standard deviation of the data qubit after the GQC. The failure occurs when the assumed error pattern is incorrect. The results confirm that our method suppresses errors more effectively than the conventional method that uses only digital information. To obtain a failure probability less than $10^{-9}$, the standard deviation should be less than 0.25 for the proposed method, whereas it needs to be less than 0.21 for the conventional method, which corresponds to the squeezing level of 9.0 dB and 10.6 dB, respectively. This improvement comes from the fact, as mentioned before, that our method can correct double errors, whereas the conventional method corrects only a single error.

{\it Concatenated code with analog information}.---In the following, we demonstrate that the proposed likelihood method improves the error tolerance on a concatenated code, which is indispensable for achieving FTQC. The use of a MLM for a concatenated code was proposed with a message-passing algorithm by Poulin~\cite{Pou}, and later Goto and Uchikawa~\cite{Goto} for Knill's $C_{4}/C_{6}$ code~\cite{Knill}. However, because previous proposals have been based on the probability of the correct decision given by Eq. (1), the error correction provides a suboptimal performance against the GQC, as shown later using a numerical calculation. 

We apply our method to the $C_{4}/C_{6}$ code modified with a message-passing algorithm proposed by Goto and Uchikawa~\cite{Goto}. The QEC in the $C_{4}/C_{6}$ code is based on quantum teleportation, where the logical qubit $\ket{\widetilde{\psi}}_{{\rm L}}$ encoded by the $C_{4}/C_{6}$ code is teleported to the fresh encoded Bell state. The quantum teleportation process refers to the outcome of the Bell measurement on the encoded qubits and determines the amount of displacement. If this feedforward is performed correctly, the error is successfully corrected. From Bell measurement, we obtain the outcomes of both bit values and deviation values for the physical GKP qubits of the encoded data qubit and encoded qubit of the encoded Bell state. Therefore, we can improve the error tolerance of the code by introducing the likelihood method to the Bell measurement (see the supplementary information for the details). 

\begin{figure}
 \includegraphics[angle=270, width=1.0\columnwidth]{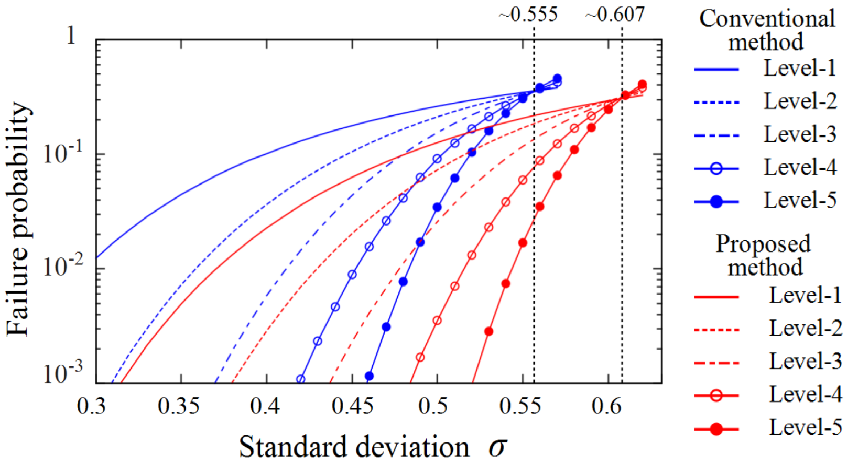}  
  \caption{\label{figc}Simulation results for the failure probabilities of the $C_{4}/C_{6}$ code using the conventional and proposed method. The failure probabilities using the conventional method (blue line) and proposed method (red line) are represented for the concatenated level-1 (solid), level-2 (dashed ), level-3 (dashed-dotted), level-4 (open circles), and level-5 (filled circles).}
\end{figure}

We simulated the quantum teleportation process for the $C_{4}/C_{6}$ code with the conventional~\cite{Goto} and proposed method using the Monte Carlo method. In this simulation, it is assumed that the encoded data qubit and encoded Bell state are prepared perfectly, and the variance of the GKP qubits of the encoded data qubit $\sigma^2$ increases only by the GQC. In Fig.\ref{figc}, the failure probabilities up to level-5 of the concatenation are plotted as a function of the data qubit's deviation. The results confirm that our method suppresses errors more effectively than the conventional method. It is also remarkable that our method achieves the hashing bound of the standard deviation for the quantum capacity of the GQC $\sim$ 0.607, which corresponds to the squeezing level of  1.3 dB and has been conjectured to be an attainable value using the optimal method~\cite{GKP,Harri}. The quantum capacity is defined as the supremum of all achievable rates at which quantum information can be transmitted over the quantum channel and the hashing bound of the standard deviation is the maximum value of the condition that yields the non-zero positive quantum capacity. By contrast, the concatenated code with only digital information achieves the hashing bound $\sim$ 0.555~\cite{GKP,Harri}, which corresponds to the squeezing level of 2.1 dB. This fact shows our method can lead to reduce the squeezing level required for FTQC.

{\it Conclusion}.---We proposed a MLM which used not only digital information but also analog information for an efficient QEC based on GKP qubits. Numerical results showed our method improved the QEC performance for the three-qubit bit-flip code and concatenated codes. In particular, we provide the first method to achieve the hashing bound for the quantum capacity of the GQC. 

Furthermore, our method can be also applied to various other codes~\cite{Kitaev2, Rau, Bom,Ste,Bac}. Therefore, the squeezing level required for FTQC with a non-concatenated code such as surface code which is used to implement topological QC~\cite{Kitaev2,Rau} can be reduced using our method~\cite{Cap2}.

Although several methods to implement GKP qubits have been proposed~\cite{Vas, Ter, Meni2, Tra, Pet, Pir, Bar} and the achievable squeezing level of a SV state is 15 dB~\cite{Vah}, it is still difficult to experimentally generate GKP qubits with the squeezing level required for FTQC~\cite{Cap3}. Our method can alleviate this requirement, and will encourage experimental developments.\\

This work was funded by ImPACT Program of Council for Science, Technology and Innovation.

\section*{Appendix A:  Three-qubit bit-flip code}
In this section, we explain how the deviation of the physical GKP qubits is projected onto the deviation of the ancillae. Fig.\ref{sfiga} shows a quantum circuit for the QEC with the three-qubit bit-flip code. This circuit looks almost the same as the circuit for DV apart from the third ancilla qubit. However, the error identification for the GKP qubits is substantially different from that for DV-QEC. In this circuit, the sum of deviations of the physical GKP qubits $i$ and $i+1$ ($i = 1, 2$) are projected onto the ancilla $i$. The deviation of the physical GKP qubit 3 is projected onto ancilla 3. First, a single logical qubit $\ket{\widetilde{\psi}}_{{\rm L}}$ is prepared by two controlled-not (CNOT) gates acting on the data qubit $\ket{\widetilde{\psi}}_{\rm 1}$ = $\alpha \ket{\widetilde{0}}_{\rm 1}$ +$\beta \ket{\widetilde{1}}_{\rm 1}$ and two ancillae $\ket{\widetilde{0}}_{i}$ ( $i$ = 2, 3). The CNOT gate, which corresponds to the operator exp(-$i\hat{q}_{\rm L}\hat{p}_{\rm A}$), transforms
\begin{align*}
\hat{q}_{\rm L} \to   \hat{q}_{\rm L}, \ \ 
\hat{p}_{\rm L} \to  \hat{p}_{\rm L} - \hat{p}_{\rm A} ,  \nonumber \\
\hat{q}_{\rm A} \to   \hat{q}_{\rm A} +  \hat{q}_{\rm L}, \ \
\hat{p}_{\rm A} \to  \hat{p}_{\rm A},
\tag{A1}
\end{align*}
where $\hat{q_{\rm L}}$ ($\hat{q_{\rm A}}$) and $\hat{p_{\rm L}}$ ($\hat{p_{\rm A}}$) are the $q$ and $p$ quadrature operators of the logical (ancilla) qubit, respectively. Then, the GQC displaces the $q$ and $p$ quadratures randomly and independently, and increases the variance of the three physical GKP qubits. After the GQC, the bit-flip error correction is implemented using the three ancillae $\ket{\widetilde{0}}_{{\rm A}j}$ ($j$=1, 2) and $\ket{\widetilde{+}}_{{\rm A}3}$. Before the CNOT gates in the error correction circuit, the true deviation values of the physical GKP qubits and ancillae in $q$ quadrature, which obey Gaussian distribution with mean zero, are denoted by $\overline{\Delta}_{j}$ and $\overline{\Delta}_{{\rm A}j}$ ($j$=1, 2, 3), respectively. For simplicity, because the ancilla qubits are fresh, we assume that the initial variance is much smaller than that of the physical qubits of the logical qubit. Then, the CNOT gates change the true deviation values of three ancillae $\overline{\Delta}_{{\rm A}j}$ in $q$ quadrature as follows:
\begin{figure}[t]
    \includegraphics[angle=270, width=1.0\columnwidth]{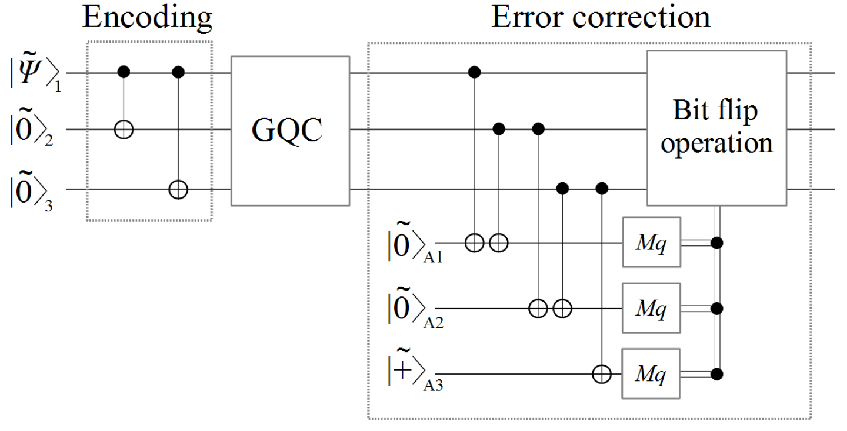}
    \caption{\label{sfiga}A quantum circuit of the QEC for the three-qubit bit-flip code with GKP qubits using the proposed method. The data qubit $\ket{\widetilde{\psi}}_{\rm 1}$ and two GKP qubits $\ket{\widetilde{0}}_{\rm 2}$ and $\ket{\widetilde{0}}_{\rm 3}$ encode a single logical qubit. $\ket{\widetilde{0}}_{\rm A1}, \ket{\widetilde{0}}_{\rm A2}, {\rm and} \ket{\widetilde{+}}_{\rm A3}$ denote ancilla qubits for the QEC. The GQC and $M_{q}$ denote the GQC and measurements of ancillae in $q$ quadrature, respectively.}   
\end{figure}
\begin{align*}
\overline{\Delta}_{{\rm A1}}  \to   \overline{\Delta}_{{\rm A1}} + \overline{\Delta}_{{\rm 1}} + \overline{\Delta}_{{\rm 2}}= \overline{\Delta}_{{\rm 1}} + \overline{\Delta}_{{\rm 2}}\ \nonumber ,\\
 \overline{\Delta}_{{\rm A2}}  \to   \overline{\Delta}_{{\rm A2}} + \overline{\Delta}_{{\rm 2}} + \overline{\Delta}_{{\rm 3}}=\overline{\Delta}_{{\rm 2}} + \overline{\Delta}_{{\rm 3}}\ \nonumber , \\
 \overline{\Delta}_{{\rm A3}} \to  \overline{\Delta}_{{\rm A3}} + \overline{\Delta}_{{\rm 3}} = \overline{\Delta}_{{\rm 3}} .
 \tag{A2}
\end{align*}
Therefore, the sum of deviations of the physical GKP qubits $i$ and $i+1$ ($i = 1, 2)$ are projected onto the ancilla $i$. The deviation of physical GKP qubit 3 is projected onto ancilla 3.

\section*{Appendix B: $C_{4}/C_{6}$ code}
\begin{figure}[b]
    \includegraphics[angle=270, width=1.0\columnwidth]{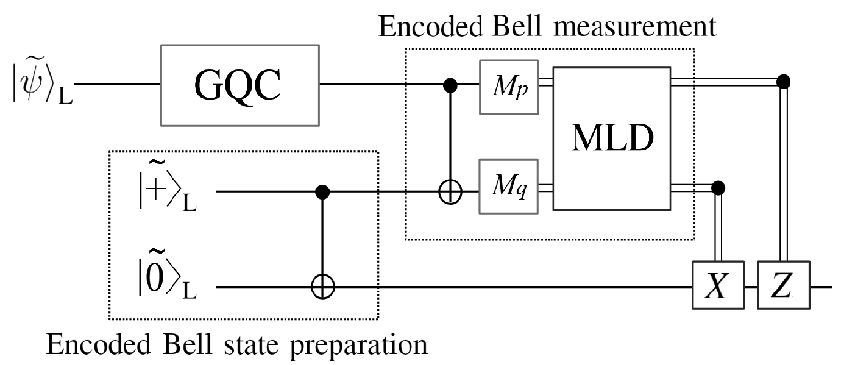}  
    \caption{\label{sfigb}Error correction by quantum teleportation. The encoded data qubit $\ket{\widetilde{\psi}}_{\rm L}$, two encoded qubits $\ket{\widetilde{+}}_{\rm L}$, and $\ket{\widetilde{0}}_{\rm L}$ are encoded by $C_{4}/C_{6}$ code. GQC and  MLD denote the GQC and a maximum-likelihood decision, respectively.}
\end{figure}

The error correction in the $C_{4}/C_{6}$ code is based on quantum teleportation, where the logical qubit $\ket{\widetilde{\psi}}_{{\rm L}}$ encoded by the $C_{4}/C_{6}$ code is teleported to the fresh encoded Bell state, as shown in Fig.\ref{sfigb}. The quantum teleportation process refers to the outcomes $M_{p}$ and $M_{q}$ of the Bell measurement on the encoded qubits, and determines the amount of displacement. We obtain the Bell measurement outcomes of bit values $m_{pi}$ and $m_{qi}$ for the $i$-th physical GKP qubit of the encoded data qubit and encoded qubit of the encoded Bell state, respectively. In addition to bit values, we also obtain deviation values $\Delta_{p{\rm m}i}$ and $\Delta_{q{\rm m}i}$ for the $i$-th physical GKP qubit. Therefore, the proposed likelihood method can improve the error tolerance of the Bell measurement.

As a simple example to explain our method for the Bell measurement, we describe the level-1 $C_{4}/C_{6}$ code, that is, the $C_{4}$ code. The $C_{4}$ code is the $[[\rm{4,2,2}]]$ code and consists of four physical GKP qubits to encode a level-1 qubit pair; thus, it is not the error-correcting code but the error-detecting code in the conventional method. The logical bit value of the $C_{4}$ code is $k$ (=0,1) when the bit value of the level-1 qubit pair is ($k$,0) or ($k$,1), that is, the bit value of the first qubit $k$ defines a logical bit value of a qubit pair. As the parity check of the $Z$ operator for the first and second qubits $ZIZI$ and $IIZZ$ indicates, the bit value of the level-1 qubit pair (0,0) corresponds to the bit value of the physical GKP qubits $(m_{q{\rm1}}, m_{q{\rm 2}}, m_{q{\rm 3}}, m_{q{\rm 4}})$ = (0,0,0,0) or (1,1,1,1)~\cite{Knill}. The bit values of the pairs (0,1), (1,0), and (1,1) correspond to the bit values of the physical GKP qubits (0,1,0,1) or (1,0,1,0), (0,0,1,1) or (1,1,0,0), and (0,1,1,0) or (1,0,0,1), respectively. Therefore, if the measurement outcome of the physical GKP qubits is (0,0,1,0) for the $Z$ basis, then we consider two error patterns, assuming the level-1 qubit pair (0,0). The first pattern is a single error on the physical qubit 3 and the second pattern is the triple errors on the physical qubits 1, 2, and 4. We then calculate the likelihood for the level-1 qubit pair (0,0) $F_{0,0}$ as
\begin{widetext}
\begin{equation*}
F_{0 ,0} = f(\sqrt{\pi}-|\Delta_{q{\rm m1}}|)f(\sqrt{\pi}-|\Delta_{q{\rm m2}}|)f(\Delta_{q{\rm m3}})f(\sqrt{\pi}-|\Delta_{q{\rm m4}}|)
                 + f(\Delta_{q{\rm m1}})f(\Delta_{q{\rm m2}})f(\sqrt{\pi}-|\Delta_{q{\rm m3}}|)f(\Delta_{q{\rm m4}}).\tag{B1}
\end{equation*}
\end{widetext} 
We similarly calculate the $ F_{0,1}, F_{1,0},$ and $F_{1,1}$ likelihood for the bit value of qubit pairs (0,1), (1,0), and (1,1). Finally, we determine the level-1 logical bit value for the $Z$ basis by comparing $F_{0,0}+F_{0,1}$ with $F_{1,0}+F_{1,1}$, which refer to the likelihood functions for the logical bit values zero and one, respectively. If $F_{0,0}+F_{0,1} > F_{1,0}+F_{1,1}$, then we determine that the level-1 logical bit value for the $Z$ basis is zero, and vice versa. The level-1 logical bit value for the $X$ basis can be determined by the parity check of the $X$ operator for the first and second qubits $XXII$ and $IXIX$ in a similar manner. In the conventional likelihood method~\cite{Pou,Goto} $F_{0,0}$, $F_{0,1}$, $F_{1,0}$, and $F_{1,1}$ are given by the same joint probability
\begin{equation*}
p_{\rm corr}^{3} (1-p_{\rm corr}) + p_{\rm corr}(1-p_{\rm corr})^{3},\tag{B2}
\end{equation*}
where the probability $p_{corr}$ is defined by Eq. (1) in the main text.
Because $F_{0,0}+F_{0,1} = F_{1,0}+F_{1,1}$, the $C_{4}$ code is not error-correcting code but error-detecting code in the conventional method, whereas it is the error-correcting code in our method. For higher levels of concatenation, the likelihood for the level-$l$ ($l\geqq2$) bit value can be calculated by the likelihood for the level-$(l-1)$ bit value in a similar manner.


\begin{thebibliography}{00}
\bibitem{Cap1}
In such a use of analog information~\cite{Dan}, qubit strings need to be measured by a readout apparatus to estimate a channel noise to obtain a likelihood using statistical approach. By contrast, a likelihood is obtained using only the measurement outcome of a single qubit in our method.
\bibitem{Cap2}
Our method can also improve the QEC performance with the non-concatenated code such as the three-qubit flip code and the $C_{4}$ code, and can straightforwardly apply to a surface code~\cite{Rau}. By contrast, previous proposals~\cite{Pou, Goto} improve the QEC performance with only the concatenated code, since the improvement results from the message-passing algorithm.
\bibitem{Cap3}
A resent proposal~\cite{Ter} exists to prepare a good GKP qubit using an interaction between a SV state and circuit QED in a current experimental set-up, and the achievable squeezing level of a squeezed vacuum state is 15 dB~\cite{Vah}. However, the process to prepare the GKP qubit, such as an interaction between a SV state and circuit QED, might decrease the squeezing level of a SV state.
\end{thebibliography}
\end{document}